\documentclass[%
superscriptaddress,
onecolumn,
10pt,
longbibliography,
amsmath,amssymb,
aps,
pra,
floatfix
]{revtex4-2}
\usepackage{graphicx}
\usepackage{dcolumn}
\usepackage{bm}
\usepackage{hyperref}
\usepackage[mathlines]{lineno}
\usepackage[T1]{fontenc}
\usepackage{enumitem}

\setcitestyle{super}

\usepackage{xcolor}
\usepackage[normalem]{ulem}

\begin{document}
\title{Data Needs and Challenges for Quantum Dot Devices Automation}

\author{Justyna P. Zwolak}
\email{jpzwolak@nist.gov}
\affiliation{National Institute of Standards and Technology, Gaithersburg, MD 20899, USA}
\affiliation{Joint Center for Quantum Information and Computer Science,
University of Maryland, College Park, MD 20742, USA}
\affiliation{Department of Physics, University of Maryland, College Park, MD 20742, USA}

\author{Jacob M. Taylor}
\email{jacob.taylor@nist.gov}
\affiliation{National Institute of Standards and Technology, Gaithersburg, MD 20899, USA}
\affiliation{Joint Center for Quantum Information and Computer Science,
University of Maryland, College Park, MD 20742, USA}
\affiliation{Joint Quantum Institute, National Institute of Standards and Technology and University of Maryland, Gaithersburg, MD 20899, USA}

\author{Reed Andrews}
\affiliation{HRL Laboratories, LLC, 3011 Malibu Canyon Road, Malibu, CA 90265, USA}%
\author{Jared Benson}
\affiliation{Department of Physics, University of Wisconsin-Madison, Madison, WI 53706, USA}%
\author{Garnett Bryant}
\affiliation{National Institute of Standards and Technology, Gaithersburg, MD 20899, USA}%
\author{Donovan Buterakos}
\affiliation{Joint Center for Quantum Information and Computer Science,
University of Maryland, College Park, MD 20742, USA}
\author{Anasua Chatterjee}
\affiliation{Center for Quantum Devices, Niels Bohr Institute, University of Copenhagen, Copenhagen 2100, Denmark}%
\author{Sankar Das Sarma}
\affiliation{Department of Physics, University of Maryland, College Park, MD 20742, USA}
\affiliation{Joint Quantum Institute, National Institute of Standards and Technology and University of Maryland, Gaithersburg, MD 20899, USA}
\affiliation{Condensed Matter Theory Center, University of Maryland, College Park, MD 20742, USA} 
\author{Mark A. Eriksson}
\affiliation{Department of Physics, University of Wisconsin-Madison, Madison, WI 53706, USA} 
\author{Eli\v{s}ka Greplov\'a}
\affiliation{Kavli Institute of Nanoscience, Delft University of Technology, Lorentzweg 1, 2628 CJ Delft, The Netherlands}%
\author{Michael J. Gullans}
\affiliation{National Institute of Standards and Technology, Gaithersburg, MD 20899, USA}
\affiliation{Joint Center for Quantum Information and Computer Science,
University of Maryland, College Park, MD 20742, USA}
\affiliation{Department of Physics, University of Maryland, College Park, MD 20742, USA}
\author{Fabian Hader}
\affiliation{Central Institute of Engineering, Electronics and Analytics ZEA-2 – Electronic Systems, Forschungszentrum Jülich GmbH, 52425 Jülich, Germany}
\author{Tyler J. Kovach}
\affiliation{Department of Physics, University of Wisconsin-Madison, Madison, WI 53706, USA}
\author{Pranav S. Mundada}
\affiliation{Q-CTRL, Santa Monica, CA 90401, USA}
\author{Mick Ramsey}
\affiliation{Intel Components Research, Intel Corporation, Hillsboro, Oregon 97124, USA}
\author{Torbjoern Rasmussen}
\affiliation{Center for Quantum Devices, Niels Bohr Institute, University of Copenhagen, Copenhagen 2100, Denmark}%
\author{Brandon Severin}
\affiliation{Department of Materials, University of Oxford, Parks Road, Oxford, OX1 3PH, UK}%
\affiliation{School of Electrical Engineering and Telecommunications, The University of New South Wales, Sydney, NSW 2052, Australia}
\author{Anthony Sigillito}
\affiliation{Department of Electrical and Systems Engineering, University of Pennsylvania, Philadelphia, PA 19104, USA}%
\author{Brennan Undseth}
\affiliation{Kavli Institute of Nanoscience, Delft University of Technology, Lorentzweg 1, 2628 CJ Delft, The Netherlands}%
\affiliation{QuTech, Delft University of Technology, Lorentzweg 1, 2628 CJ Delft, The Netherlands}%
\author{Brian Weber}
\affiliation{National Institute of Standards and Technology, Gaithersburg, MD 20899, USA}
\affiliation{Intelligent Geometries, LLC, Lake Frederick, VA 22630, USA}

\date{\today}
\begin{abstract}
Gate-defined quantum dots are a promising candidate system for realizing scalable, coupled qubit systems and serving as a fundamental building block for quantum computers.
However, present-day quantum dot devices suffer from imperfections that must be accounted for, which hinders the characterization, tuning, and operation process.
Moreover, with an increasing number of quantum dot qubits, the relevant parameter space grows sufficiently to make heuristic control infeasible. 
Thus, it is imperative that reliable and scalable autonomous tuning approaches are developed.
This meeting report outlines current challenges in automating quantum dot device tuning and operation with a particular focus on datasets, benchmarking, and standardization.
We also present insights and ideas put forward by the quantum dot community on how to overcome them. 
We aim to provide guidance and inspiration to researchers invested in automation efforts.
\end{abstract}

\maketitle

\section{Background and motivation}\label{sec:motivation}
Gate-defined semiconductor quantum dots are a candidate system to realize scalable, coupled qubit systems that serve as a fundamental building block for quantum computers~\cite{Burkard21-SSQ, Chatterjee21-SQP}. 
Their potential for leveraging the semiconductor industry's materials science and fabrication techniques, while promising, remains hard to realize at scale~\cite{Neyens24-PQW}. 
Specifically, current quantum dot devices suffer from variations in the device dimensions and properties, as well as defects.
The corresponding disorder in the electronic landscape throughout a device can be overcome by using local gate voltages to tune the device into an operation regime for quantum computing. 
However, even tuning a double quantum dot device constitutes a nontrivial task, with each dot typically being controlled by at least three metallic gates, each of which influences the number of electrons in the dot, the tunnel coupling to the outer leads, and the interdot tunnel coupling, which are critical parameters for qubit operation.

The current practice of characterizing and tuning quantum dots for qubit operation either manually or using script-based methods is a relatively time-consuming procedure that is inherently impractical for scaling up and applications.
Moreover, with an increasing number of quantum dot qubits, the relevant parameter space grows sufficiently to make heuristic control unfeasible. 
To overcome the limitations of human-driven experimental control, researchers working with semiconducting quantum dot devices have put considerable effort into automating device control and characterization~\cite{Zwolak21-AAQ}. 

To date, several automated tuning algorithms for single- and double-quantum-dot devices have been proposed and demonstrated.
While initially all efforts focused on developing in-house script-based algorithms that were tailored to a particular device, more recently members of the community have begun to take advantage of the data analysis tools provided by the field of artificial intelligence and, more specifically, supervised and unsupervised machine learning~\cite{Zwolak21-AAQ}. 
When provided with proper training data, machine-learning-driven and machine-learning-enhanced methods have the flexibility of being applicable to various devices with minimal to no adjustments or retraining~\cite{Zubchenko24-ABQ}. 
Moreover, by learning the governing rules and dynamics directly from the data, such algorithms may be less susceptible to programming errors. 

However, machine learning models typically require large, labeled datasets for training, validation, and benchmarking and often lack information about the reliability of the output. 
Moreover, since the application of machine learning to quantum dot tuning, characterization, and control is a relatively new field of research, it lacks standardized measures of success. 
The reported success rates vary significantly in both the level and meaning of the reported performance statistics, making it hard (if not impossible) to benchmark the proposed techniques against more traditional tuning approaches or against one another~\cite{Zwolak21-AAQ}.
The time is thus ripe to discuss the broadly defined needs and potential next steps in the field of quantum dot device automation that would enable substantive progress. 
 
A simple but crucial component of success for the field will be to solidify key metrics of performance as well as establish standard datasets that can be used to assess the performance of the newly proposed methods and algorithms. 
Among the simple metrics that have been used to date are \textit{state identification accuracy}, defined as the probability of a classifier identifying the right device topology, and \textit{tuning success}, defined as the probability of the navigation algorithm getting to the right region of parameter space. 
However, more specialized metrics, and associated datasets, will be necessary to leverage automation algorithms most effectively. 

When it comes to datasets, most characterization and tuning efforts undertaken to date rely on datasets that consist of either simulated data (which may lack important features representing real-world noise and imperfections) or manually labeled experimental data (which might be subject to qualitative and/or erroneous classification). 
Moreover, with a few exceptions, experimental datasets are not made publicly available. 
At the same time, systematic benchmarking of tuning methods on standardized datasets, analogous to the MNIST~\cite{LeCun98-MNIST} or CIFAR~\cite{Krizhevsky09-CIFAR} datasets, is a crucial next step on the path to developing reliable and scalable autotuning algorithms for quantum dot devices.

The goal of the \textit{Workshop on Advances in Automation of Quantum Dot Devices Control} was to serve as a starting point for discussions about the community's needs and interests~\cite{AQD2023}.
For two days stakeholders from industry, academia, and the government interested in research and development of semiconductor quantum computing technologies discussed methods of collaboration and future roadmap development of methods for tuning large-scale devices.
Topics discussed during the meeting included:
\begin{itemize}[topsep=3pt, itemsep=-2pt]
    \item opportunities for research and development of tuning, characterization, and control methods for semiconductor quantum dot devices;
    \item identifying barriers to near-term and future applications of the autotuning methods;
    \item determining key performance metrics for the various aspects of the tuning, characterizing, and controlling of quantum dot devices; and
    \item the need for facilitating interaction and collaboration between the stakeholders to build a large open-access database of experimental and simulated data for benchmarking new autotuning algorithms. 
\end{itemize}

\begin{figure*}[t]
    \centering
    \includegraphics[width=0.95\textwidth]{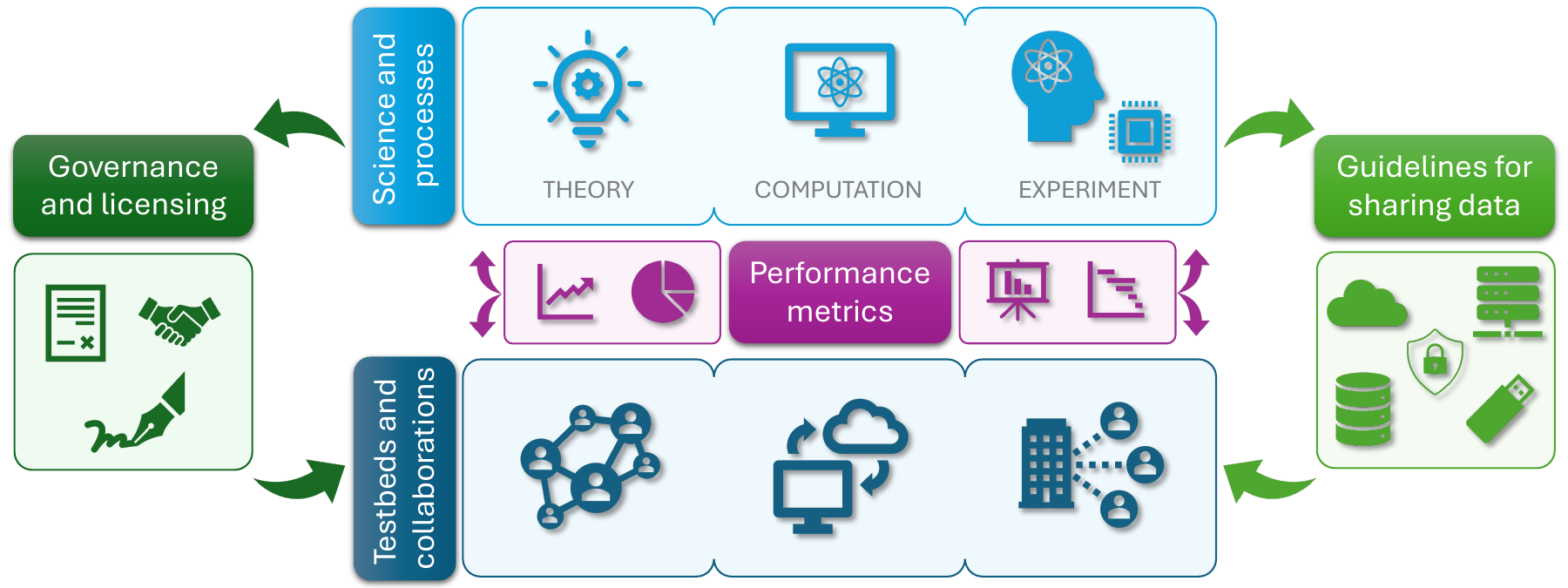}
    \caption{ 
    {\bf Workshop thematic map.} 
    The workshop discussions revolved around five main themes: science and policy challenges; guidelines for sharing data; developing performance metrics; establishment of testbed and collaborations; and governance and licensing.
    These themes are connected to allow the flow of data and information.
    The end result is an infrastructure for facilitating the development, benchmarking, and standardization of the quantum dot automation methods, where the science is the driver (top row), the testbeds and collaborations are critical (bottom row), and the governance (left box) and the guidelines (right box) take their inputs from the science and provide useful help to the testbed and collaborations.
    For example, theory, computation, and experiment provide a description of needs that can be met through testbeds and collaboration, the latter two of which are facilitated by governance and guidelines.
    }
    \label{fig:framework}
\end{figure*}

The workshop discussions, guided by a set of questions first put forward in a Federal Register Notice published on April 13, 2023~\cite{FRN23}, revolved around five main themes, which we depict in Fig.~\ref{fig:framework} present in the following sections.
In terms of datasets and benchmarking, the workshop participants were asked to identify public or restricted-use datasets related to the various phases of tuning semiconductor quantum dot devices that are available for training and benchmarking new artificial intelligence models or to test hypotheses using data mining/machine learning methods. 
They were asked to consider the work researchers need to do to access, and then explore the quality of, existing datasets before conducting research with them and to identify what aspects of this work could be reduced or conducted just once so that future researchers can reduce the time needed to complete a research project.
The workshop participants were also asked to consider best practices for creating new datasets or linking existing datasets and sharing them with researchers while adhering to local, State, and Federal laws, as well as barriers and limitations that currently exist to data sharing.
Finally, they were asked to think about what role the National Institute of Standards and Technology (NIST) can play in developing infrastructure that supports the use of large-scale datasets for research on tuning quantum dot devices.

Focusing on automation, the workshop participants were asked to consider to what extent existing datasets capture enough information to address research related to all aspects of tuning quantum dot devices, to describe the research needs that are not being met by the currently available datasets, and to identify additional data that should be collected to address these research questions.
They were also asked to think about whether existing datasets, both simulated and acquired experimentally, contain data that are valuable for researchers and are of sufficient quality that research could be conducted with a high amount of rigor.
Part of the discussions revolved around the promising approaches to testing and improving the validity of performance metrics within large datasets, especially those datasets that consist of experimental data that do not come with ground truth labels. 

In addition to the overall comprehensive summary, each thematic section of this report concludes with a set of core recommendations for possible next steps for advancing automation of the quantum dot tuning and characterization.
These recommendations do not represent an endorsement by or an official position of NIST or any other agency participating in the workshop.
Rather, they should be understood as summarizing the plurality of views from the workshop.      

The core recommendations from the workshop are given here, and substantiated in the following section: 
\begin{itemize}[topsep=3pt, itemsep=0pt]
\item[A.] Science and process challenges
\begin{enumerate}[topsep=0pt, itemsep=2pt]
    \item The community must directly investigate methods and approaches for rapidly tuning increasingly large arrays in one and two dimensions. 
    To be robust, this examination must consider the meaning and intent of success and failure of subroutines and subprocesses in the tuning method, enabling learning for future experiments when the tuning process as a whole fails.

    \item The community must work together to ensure that measurement approaches and techniques enable effective modeling in a standardized form.
    This should result in a series of interrelated, physics-based models, such as capacitance relationships and multiband Hubbard models.
\end{enumerate}

\item[B.] Guidelines for sharing data
\begin{enumerate}[topsep=0pt, itemsep=2pt]
    \item NIST could work with the community to define the ontology and core guidelines under the FAIR (findable, accessible, interoperable, and reusable) principles, along with a living process for improving and revising these guidelines, to enable data sharing across the community.
    
    \item These guidelines would ideally be accompanied by a community-enabling data-sharing service with core example datasets and with an IP policy tailored to each of the datasets that prioritizes sharing.
    
    \item The community should work to determine additional sharing needs and the potential for a sharing collaborative network that enables members to share data in a protected and secure manner.
    
    \item As the community works on the above topics, it can benefit substantially by developing a few example datasets to test these guidelines and sharing principles.
\end{enumerate}

\item[C.] Performance metrics
\begin{enumerate}[topsep=0pt, itemsep=2pt]
    \item Metrics should be developed in the context of specific tuning or calibration routines and should be referenceable between different devices and different groups by using standardized units where possible as well as standardized practices, such as the use of virtual gates. 

    \item Given that different routines are at different levels of common usage and similarity, e.g., some phases of device tuneup might be nearly ready for standardization while others are in an early development stage, each should be treated according to its maturity and breadth acceptance and/or use.
   
    \item Labeling of data should, where possible, be complemented with all available physics or materials knowledge of the device(s) and design(s) used in the measurement of the data to ensure that quantitative understanding and systematic differences are properly integrated into cross-device characterization. This device-level information should be included in the metadata of the dataset.
\end{enumerate}

\item[D.] Testbeds and collaborations
\begin{enumerate}[topsep=0pt, itemsep=2pt]
    \item The community should develop and refine a prioritized list of physical and engineering principles to center collaboration activities and identify needs for testbeds or other facilities. This could follow the different stages of tuneup from device characterization all the way to high-fidelity, few-qubit operation.    
    \item Physical testbeds can be effectively complemented with model-based test systems. Such virtual testbeds should have ease of use for a wider community, and require experimental validation and verification.
    \item Testbeds and other systems using standardized approaches can provide the foundation for a reference implementation of a complete, scaling quantum dot system.
\end{enumerate}

\item[E.] Governance and licensing
\begin{enumerate}[topsep=0pt, itemsep=2pt]
    \item The community should establish a working group to oversee the governance and licensing of datasets and to inform database maintainers' choices. 
    NIST could act as a convener of this working group. 
\end{enumerate}
\end{itemize}

\section{Automation challenges}
\label{sec:challenges}

\subsection{Science and process challenges}
At a high level, quantum dot systems have a variety of engineering and science challenges. 
However, the largest, pressing challenge is developing a reliable path towards large arrays, in one or two dimensions (or other geometries with scaling at least proportional to the linear size of the system), of quantum dots that are tunable into the quantum coherent, few-electron regime~\cite{Li18-CNQ}.

An example of the challenge ahead was characterized by the difficulties of direct emulation of the Hubbard model. 
In 2019, Vandersypen and co-authors demonstrated Nagaoka ferromagnetism in a small quantum dot array with four dots~\cite{Dehollain19-FQD}. 
This required building a Hamiltonian-level understanding of the device and maintaining the device in a narrow regime of Hamiltonian parameters throughout the experiment. 
The time to tune up, and retune, the device becomes a potential limiting factor for scaling to larger numbers of qubits (or, in this case, Hubbard model sites). 

This speaks to the underlying need for rapid characterization and calibration processes that can be effectively interleaved with scientific data gathering, and that can most efficiently move the setup towards the desired regime. 
It also showcases the necessity of reducing the cost of tuning up additional qubits, so that making larger arrays does not take polynomially more time (in the absence of low-frequency noise) or, effectively, forever in the presence of low-frequency noise. 

At the same time, characterization and calibration rely upon an understanding of the materials, the device design and geometry, the experimental control systems, and the experimental measurement systems. 
Thus, a related sub-challenge is providing physics- and materials-level knowledge via typical calibration and characterization routines. 
For example, \textit{pinch-off} -- the nominal voltage below which measurable current will no longer flow even quantum mechanically -- is a key characteristic of a gate in gated devices. 
Connecting the voltage of a gate to the local chemical potential for the Fermi sea (i.e., setting a lever arm) is an essential calibration step but today's data gathering and storage typically uses reported values (millivolts applied) rather than derived values (microelectron volts of chemical potential). 
Agreeing upon methods of estimating and storing the connection between the device and derived values enables comparison across devices and the creation of more standardized routines and more standardized expected ranges of parameters, which can make datasets interoperable.

A multi-physics modeling endeavor provides one path that could lead to a robust and reliable approach to scaling. 
This approach uses a series of interrelated models yielding a hierarchical set of relationships, from basic semiconductor calibration tests to capacitance models and detector models, to a multiband Hubbard model description of the few electron, many dot regime. 
In many respects, each model used in this approach is partially defined by the calibration or characterization experiment it seeks to describe. 
Validating related models to create the larger, multi-physics effort will necessarily require back and forth between experimental, theoretical, and software and hardware engineering efforts.

The key benefit of this hierarchical set of models is the ability to quantify the shared assumptions necessary to export and import calibration and characterization processes and routines. 
The process of developing this set of models also better informs the meaning and intent of the \textit{failure} of a particular measurement or step. 
Ideally, it would allow for shared techniques and agreements about documentary standards such that multiple groups and/or companies would be able to implement reference examples, as well as work on a dataset curated for benchmarking such reference implementations.

Our specific recommendations:
\begin{enumerate}[topsep=3pt, itemsep=0pt]
    \item The community must directly investigate methods and approaches for rapidly tuning increasingly large arrays in one and two dimensions. 
    To be robust, this examination must consider the meaning and intent of success and failure of subroutines and subprocesses in the tuning method, enabling learning for future experiments when the tuning process as a whole fails.
    
    \item The community must work together to ensure that measurement approaches and techniques enable effective modeling in a standardized form.
    This should result in a series of interrelated, physics-based models, such as capacitance relationships and multiband Hubbard models.
\end{enumerate}

\subsection{Guidelines for sharing data}
Advances in the automation of the characterization and tuning of quantum dot devices cannot happen without the joint effort of the whole community.
The development of specialized automation techniques, both machine-learning- and non-machine-learning-driven, is at present stymied by the lack of high-quality relevant datasets.
Although some groups already share their data using general-purpose open repositories such as Zenodo~\cite{Zenodo} or Open Science Framework~\cite{OSF}, such data is typically unlabeled and limited in scope to include only the \textit{good} data that was measured to create the final version of the paper.
While a lot of things can go wrong with measurement, the negative counter-examples representing \textit{bad} data (e.g., poor quality measurement, measurements over incorrectly set parameters) or failure modes are rarely made available. 
This makes it challenging to learn from mistakes and results in multiple groups reproducing the same errors. 
Moreover, in some groups, it is still a fairly common practice for researchers to store data on private repositories and to make data available only upon ``reasonable request,'' or to not share it at all. 

While the initial data-sharing efforts indicate that there is interest in both sharing the already acquired data and measuring new data, as a community, we are far away from a standardized database that could be used to develop and benchmark new and existing tuning methods.
One of the main obstacles on the path to establishing a comprehensive and holistic database of quantum dot data is the lack of guidance on how to facilitate such a process.
A number of factors important for reusability and reproducibility need to be considered to ensure that the resulting database meets the needs of the community.
These include the structure of the individual data files, comprehensive data documentation, and the inclusion of additional information related to the type of device used in the experiment, data acquisition tools, pre-processing techniques, etc. (contextual metadata). 

A primary barrier that experimentalists in the community face is that the development of control software, and even changing existing software requires significant effort.
Specifically, the software tools used to control and measure the experimental devices vary between groups, from software available commercially such as Labber~\cite{Labber}, to open-access data acquisition frameworks such as QCoDeS~\cite{QCoDeS}, to custom, in-house build packages.
While many software companies enable certain kinds of standardized measurements, routines, and data processing, the resulting data file structures are not consistent between groups. 
Even labeling the various gates in a device (e.g., plungers vs. barriers) is at present a local convention and may lead to confusion when different groups try to use the same data for testing. 
Any centralized scheme should consider that a data-sharing tool should be almost automatic for researchers to consider transitioning from their currently custom and tailored data acquisition and storage tools to community standards; only this can ensure widespread adoption. 
One solution that could be implemented is to embed seamless and transparent data sharing within the most commonly used sofftware frameworks (QCoDeS, Labber), and to provide a module easy to integrate into others (for example, in Python and/or Matlab).

Another challenge in data sharing is the inclusion of contextual metadata necessary to facilitate reproducibility.
Even if all voltages applied to all gates are saved, this information might be insufficient to replicate the experiment. 
A lot of relevant information, such as the exact magnetic field used in experiments, the temperature in the dilution fridge, or the temperature in the room, is currently not captured by the automated measurement routines and is thus typically missing from the data.  
At the same time, a measured parameter range (e.g., two columns of voltages applied to plunger gates) and the resulting device response (e.g., measured differential conductance) are not enough to transfer the full information about the experiment.
It thus might be worthwhile to save the complete device setup and a measurement history for the experiments as part of the metadata.

Establishing a centralized space where it would be easy for everyone to upload and access the data is yet another aspect of data sharing that needs to be addressed.
First of all, the space dedicated to storing the quantum dot database should provide an easy interface to upload and access the data.
Given the rapid development of programming languages, it is also important that the data files are accompanied by minimal reproducible examples of source code to load and preview the data that includes information about the package requirements.
Finally, the data should be easily accessible while also stored securely to prevent any data corruption.

Given the above discussions, several core principles should be encouraged for standardizing data.
In addition to the high-level FAIR data principles~\cite{Wilkinson16-FAIR, GO_FAIR}, there are several additional requirements specific to experimental data:
\begin{itemize}
    \item[\textbf{{[T]}}] \textbf{Traceability} 
    \begin{itemize}[nosep]
        \item[(T1)] Clear connection between numbers in the data files and experimental parameters.
        \item[(T2)] Clear connection between the experimental parameters and the device layout (gate on a micrograph), geometry, and scale, including heterostructure information. 
        \vspace{-0.5\baselineskip}\mbox{}
    \end{itemize} 
    
    \item[\textbf{{[A]}}]\textbf{Accessibility}
    \begin{itemize}[nosep]
        \item[(A1)] Providing program files or documentation to enable others to use the data. 
        \item[(A2)] Documenting libraries reacquired for using the data, including version when applicable. 
    \end{itemize}
    
    \item[\textbf{{[S]}} ]\textbf{Standarization}
    \begin{itemize}[nosep]
        \item[(S1)] Using standard file formats for data storage.
        \item[(S2)] Using standardized units that are clearly identified for each measurement to store the data. 
        \item[(S3)] Using standardized and/or well-documented procedures.
        \item[(S4)] Using standardized measurement subsystems and/or measurement techniques.
        \vspace{-0.5\baselineskip}\mbox{}
    \end{itemize}
    
        \item[\textbf{{[C]}} ]\textbf{Contextuality}
    \begin{itemize}[nosep]
        \item[(C1)] Identifying measurement techniques and procedures used to acquire the data.
        \item[(C2)] Including metadata (e.g., magnetic field, the temperature in the dilution fridge and of the electron gas, device configuration on gates not actively measured).
        \item[(C3)] Articulating the assumptions about results from preceding measurements (what previous data does it depend upon).
        \item[(C4)] Articulating the purpose of the measurement (what information will it provide, what questions does it answer for future data).
        \item[(C5)] Linking to (future) key qubit metrics such as readout and one- and two-qubit gate fidelities, where possible.
        \vspace{-0.5\baselineskip}\mbox{}
    \end{itemize}

\end{itemize}

The core recommendations from the workshop for sharing data and building the quantum dot database are the following:
\begin{enumerate}[topsep=3pt, itemsep=0pt]
    \item NIST could work with the community to define the ontology and core guidelines under the principles of FAIR TASC, along with a living process for improving and revising these guidelines, to enable data sharing across the community.
    
    \item These guidelines would ideally be accompanied by a community-enabling data-sharing service with core example datasets and with an IP policy tailored to each of the datasets that prioritizes sharing.
    
    \item The community should work to determine additional sharing needs and the potential for a sharing collaborative network that enables members to share data in a protected and secure manner.
    
    \item As the community works on the above topics, it can benefit substantially by developing a few example datasets to test these guidelines and sharing principles.
\end{enumerate}

\subsection{Performance metrics}
A pivotal component of success for advancing the automation of quantum dot device characterization and control will be to solidify key metrics of performance and to establish standardized reference datasets that can be used to assess them. 
Defining good performance metrics is a really difficult task, especially in the earlier phases of tuning.

A few simple metrics, such as \textit{state identification accuracy} (the probability of a classifier identifying the right device topology) and \textit{tuning success} (the probability of an optimizer or other navigation algorithm getting to the right region of parameter space), have been used for the middle to later stages of tuning~\cite{Kalantre17-MLD, Zwolak20-AQD, Darulova19-ATQ, Durrer19-ATQ, Czischek21-MNA, Zwolak21-RBI, Ziegler22-TAR}.
The performance of fine-tuning and gate-tuning processes is often assessed through the final fidelity of the qubit through, e.g., randomized benchmarking.
Additional metrics for single-qubit operations could be based on other experimental measurements, such as the number of visible Rabi oscillations, where more observed oscillations indicate superior performance. 
Similar metrics, e.g., the quality of exchange oscillations, should also be developed for two-qubit gates.
However, defining more rigorous quantitative and qualitative metrics and processes to assess the expected overall performance of devices remains an open problem in the field that will require the engagement of the whole community.
Focusing on both single-qubit and two-qubit gate standardization will be necessary to avoid calibrating an array of single qubits that can not talk to each other. 
Once standardization is in place, more structured and algorithmic exploration and improvement of the tuning processes can be achieved. 

To date, there have been two approaches to testing the performance of tuning algorithms: (i) direct deployment on experimental devices~\cite{Baart16-CAT, Darulova19-ATQ, Durrer19-ATQ, Moon20-ATQ} and (ii) using simulated device data~\cite{qf-data, Zwolak18-QLD} for development and initial assessment of tuning methods followed by testing on experimentally acquired data or directly on experimental devices~\cite{Kalantre17-MLD, Lapointe-Major19-ATQ, Zwolak20-AQD, Zwolak21-RBI, Ziegler22-TAR, Czischek21-MNA, Schuff23-IPS}.
The benefit of using simulated data is access to information about what the data represents (e.g., the exact charge occupation or the state of the device) as well as the ability to facilitate a controlled study of how the various changes to the device design or the types, combinations, and prevalence of noise impact the functioning of the tuning algorithms.
At the same time, simulation tools are not equipped to cover all aspects of the tuning process and may not capture some of the realistic experimental imperfections~\cite{Darulova20-EDM, Ziegler22-TRA}.
    
While simulated data might be a good starting point for developing and testing new tuning methods, for benchmarking combining synthetic and experimental datasets might be a more appropriate approach.
However, unlike simulated data, experimental data does not come with labels, and manual data labeling is not an easy task.
It typically involves a team of annotators, ideally, experts in the subject matter, who review each data point and assign the appropriate label based on labeling guidelines and their understanding of the data.
The process can thus be very time-consuming and labor-intensive, and result in sub-optimal labels due to the complexity of the quantum dot data. 
Even the seemingly straightforward task of determining whether a given charge stability diagram represents a \textit{good double dot} can be challenging as the quality \textit{good} might have very different meanings for different domain experts.
Similarly, qubit-specific tuning may be required.
For example, tunnel rates between dots are essential for two-qubit gates single-spin qubits~\cite{veldhorst2015-Two, Watson18-TQP} and even for individual qubit operation for exchange-only~\cite{Medford13-REQ} or quantum dot hybrid qubits~\cite{Shi12-HSQ}, and the tuning of these tunnel rates interacts with the optimum choice for gate voltage control pulses.  An example of this type of interaction is visible in so-called ``fingerprint plots'' for the exchange-only qubit~\cite{Reed2016-fingerprint}.
The development of a general, systematic, unbiased, and preferably automated labeling procedure might be necessary if experimental data is to be included in a database intended for benchmarking~\cite{Weber23-PAA}.
Such effort needs to be carried out as a  collaboration between computational, theoretical, and experimental groups to ensure a satisfying performance of the resulting labeling software.

In addition to the label reliability, there is also the question about the types of labels that would be useful to the community.
Researchers have different needs for the labeling and usage of data and, depending on which tuning phase a particular dataset is intended to support, different labeling schemes might be necessary.
A lot of tuning aspects are specific to a particular experimental setup, the material used to fabricate the device, and the device design.
While in some applications it might be sufficient to know the quality \textit{score} for the data (i.e., a binary \textit{good} vs.\ \textit{bad} pinch-off curve or charge stability diagram), others might need labels that provide more detailed information such as \textit{category} (e.g., stability diagram capturing \textit{single} vs.\ \textit{double} quantum dot state; an exact \textit{charge occupation}), \textit{pixel categorization}, or perhaps a \textit{graph}. 
In addition to the main categories, secondary labels could include some kind of score metric concerning, for example,  the noise level or type or measurement visibility. 

Finally, complementary to the performance at the tuning phase level different categories of device functionality also come with performance metrics.
The success of tuning in this context will depend on the objective of a given experiment. 
Some of the categories worth considering include the following:
\begin{itemize}[topsep=3pt, itemsep=0pt]
    \item maintaining the same charge state on a quantum dot;
    
    \item stability of qubits tuned for single-and two-qubit operation within desirable fidelity range; and
    
    \item stability of device parameters such as tunnel coupling or onsite energy within a particular specification allowing to, e.g., perform interesting simulations.
    \end{itemize}
    
Standardized ways to characterize and label the experimental data are crucial for the development of reliable and scalable tuning methods.
The core recommendations from the workshop related to performance metrics are as follows:
\begin{enumerate}[topsep=3pt, itemsep=0pt]
    \item Metrics should be developed in the context of specific tuning or calibration routines and should be referenceable between different devices and different groups by using standardized units where possible as well as standardized practices, such as the use of virtual gates. 

    \item Given that different routines are at different levels of common usage and similarity, e.g., some phases of device tuneup might be nearly ready for standardization while others are in an early development stage, each should be treated according to its maturity and breadth acceptance and/or use.
   
    \item Labeling of data should, where possible, be complemented with all available physics or materials knowledge of the device(s) and design(s) used in the measurement of the data to ensure that quantitative understanding and systematic differences are properly integrated into cross-device characterization. This device-level information should be included in the metadata of the dataset.
\end{enumerate}

\subsection{Testbeds and collaborations}
For each critical calibration or characterization task, there is an opportunity to define and refine the associated task in both experiments and theory. 
This can occur in a focused manner through the use of dedicated facilities -- testbeds -- that develop, refine, and make such core techniques interoperable. 

Critically, a testbed brings with it a set of approaches that are different than typical experimental physics development. 
These include:
\begin{itemize}[topsep=3pt, itemsep=0pt]
    \item documentation and training of shared techniques and approaches to enable users to rapidly come up to speed on the system;

    \item shared code and databases to seed best practices and routines back into the community;

    \item access to state-of-the-art quantum dot device chips and instrumentation; 

    \item rapid test and characterization capabilities, to more fully explore a parameter space; and
    
    \item community input into focus and use.
\end{itemize}
We recognize that organizing a testbed facility specifically for a key physics, material science, electrical engineering, software engineering, or quantum computing goal may be both beneficial and necessary to ensure the end-to-end operation of the system. 
For example, a testbed focused on rapid material testing is likely to invest in a set of instrumentation and other equipment that makes this testing effective but would be unlikely to be suited for large-scale qubit experiments.

A key point developed in the course of discussions was the potential benefit of a virtual testbed facility, validated from experimental data but supported by theoretical model(s), such as the multiphysics model described in the ``Science and process challenges'' section above. 
This virtual facility could enable both online and offline testing of characterization routines and calibration approaches against validated models, up to and including potentially qubit physics, without requiring the underlying quantum hardware.
Such testing and comparison would in turn help identify the most promising approaches that should be further developed to establish best practices, as opposed to the current methods that remain mostly siloed to individual groups.

The virtual testbed approach could complement existing programs such as the Quantum Foundry program~\cite{QuantumFoundry} and the LPS Qubit Collaboratory~\cite{LQC}. 
We wish to emphasize that a virtual testbed is only as good as its access to and validation with experimental systems that are representative of devices of interest. 
At the same time, the ability to make these virtual systems widely accessible is clear, since software is cheaper than hardware. 

Over time, the development of testbeds can enable device standardization, which in turn will enable more effective modeling and the development of the approaches necessary to enable robust and rapid scale-up. 
Device standardization can also be complemented with routine and process standardization. 
For example, one can imagine the creation of reference implementations for the control and tuning stack, which enable research teams and engineering teams to produce a starting, functional system from which more complex future devices can be developed.

The key recommendations from this section include:
\begin{enumerate}[topsep=3pt, itemsep=0pt]
    \item The community should develop and refine a prioritized list of physical and engineering principles to center collaboration activities and identify needs for testbed or other facilities. 
    This could follow the different stages of tuneup from device characterization all the way to high-fidelity few qubit operations.
    
    \item Physical testbeds can be effectively complemented with model-based test systems.
    Such virtual testbeds should have ease of use for a wider community, and require experiment validation and verification.
    
    \item Testbeds and other systems using standardized approaches can provide the foundation for a reference implementation of a complete, scaling quantum dot system.
\end{enumerate}

\subsection{Governance and licensing}
Equally important as the data storage infrastructure is identifying an organization, or a collaboration, that will lead the efforts.
One possibility would be to establish a \textit{working group}, i.e., a small group of experts invested in the field of quantum dot characterization and tuning, who would lead the discussions and activities around building and maintaining the database.
Careful consideration needs to be given when deciding who should be the governing body for the effort and how the initial members of the working group should be determined. 
The priorities might be very different depending on the type of organization leading the effort, particularly if it is a commercial entity.
For example, a hardware-focused company might have different expectations regarding the resulting database than a more software-focused one.
They also might have very different organizational rules. 
Choosing a neutral hosting site that is DOI referenceable, such as \href{https://data.gov/}{data.gov}, could be pivotal for bridging the gap between the different institutions' capabilities and interests related to data sharing, ensuring a mutually beneficial interaction, and minimizing the risk of potential conflicts of interest.

The initial objective of the working group would be to work together as representatives of the larger community to develop a formal agreement between industry, government, and academia that would guide the collaborative efforts.
Such an agreement would settle the ground rules of engagement in place, provide a quality standard for collaborations, clarify intellectual property rights, and specify the license (or a set of licenses) under which the data is shared.
The best practices to ensure data safety, i.e., preventing accidental loss or manipulation of the data, and data security, i.e., preventing intentional theft or manipulation, also need to be a part of the agreement.
New partners interested in joining the collaboration would be required to 
review and agree to comply with the terms of such a governing document.

Creating common rules of engagement is critical given the different goals and priorities that the partners within the collaboration might potentially have for the effort. 
Thus, determining the desirable characteristics of potential members of the working group is a key part of building a successful team.
The outcome of a working group, how quickly it achieves its goals, and how inclusive it is all depend on the ground rules and attendees. 
There are several things to consider when building a working group:
\begin{itemize}[itemsep=1pt]
    \item[] \textbf{\textit{A clear vision of the goals.}} 
    The working group members need to be able to distinguish between the objectives of the working group and their personal or institutional preferences.
    \item[] \textbf{\textit{Subject matter expertise.}} 
    Choosing members solely based on their seniority within their respective institutions will not necessarily lead to making progress on the working group’s mission.
    \item[] \textbf{\textit{Communication skills.}} 
    Members of the working group need to be able to communicate effectively and without invoking authority, especially with those who might not share their perspectives.
    \item[] \textbf{\textit{Diversity and inclusion.}} 
    Harnessing the expertise, knowledge, and unique viewpoints of members from varying backgrounds to foster creativity and lead to innovative solutions.
\end{itemize}

It is important to clearly set expectations about what being a part of the working group entails and what is expected of future partners and collaborators regarding sharing versus using the data.
In the case of the industry partners, there might be legal constraints put in place by funding agencies or by the company's management preventing researchers from openly sharing the data. 
Being dependent on national and institutional priorities, certain national laboratories and government agencies might also be limited in what data can be shared given the priorities that they have for their agency.  
National (or even international) interests need to be considered in terms of sharing or creating datasets.
Moreover, partnerships with overseas institutions might be constrained due to the involvement of U.S.-based research institutions and companies with the military through funding by agencies such as the Defense Advanced Research Projects Agency or the U.S. Army DEVCOM Army Research Laboratory.
At the same time, it would not be helpful if only Europe or only the U.S. focused on these efforts -- it should be a collaborative effort where all interested parties have an opportunity to contribute. 

Another aspect critical to building a successful working group is ensuring diversity in terms of experience (graduate students, postdoctoral researchers, early career researchers, principal investigators), demographics (ethnicity, race, gender), research focus (experimentalists, theorists, computational scientists, engineers), and type of institution (government laboratories, universities, private sector).
The diversity of ideas, perspectives, and backgrounds, often results in a much more creative solving of complex problems than would be possible for individuals. 

When choosing the data-sharing platform it is important to ensure that both sides, i.e., those who contribute and those who access the database, have access to the data, making it very easy for everyone to share and reuse the data.  
An excellent example of a digital ecosystem of openly available data and supporting tools is the Community Resource for Innovation in Polymer Technology (CRIPT) platform -- a partnership between the University of Chicago, NIST, Massachusetts Institute of Technology, Dow Inc., and Citrine Informatics~\cite{CRIPT}.
CRIPT provides researchers working in polymer science and engineering with easy access to a large database of polymer data as well as a set of tools for interacting with the CRIPT platform.
The European Organization for Nuclear Research (CERN) Open Data Portal~\cite{CERN-ODP, Cowton15-ODP} and HEPdata~\cite{HEPdata, Maguire17-HEP} which gives access to data produced through the research performed at CERN are two examples of data-sharing platforms from high energy physics.
The Phase Field Community Hub~\cite{PFHub} provides a space where the phase field practitioners and code developers can share code and compare code output data using a standard set of metrics.
These platforms could serve as examples of successfully deployed data- and code-sharing environments.

The last aspect of establishing a database that the working group needs to consider is choosing a license (or a set of licenses) as well as clarifying the intellectual property rights governing the use of the database.
The database terms of use need to be transparent about the data usage to ensure that users contributing new data know what is happening with the data that is being shared.
It should explicitly state the users' obligations as well as prohibited uses.
The expectations and attributions need to be clearly outlined, especially regarding new ideas coming out of the shared data.
It needs to be made explicit what will happen when the data is used for research and academic purposes to elevate the concerns with sharing data and make it clear what the benefits of sharing data are. 

In the end, an overall understanding has to be reached about who should be the driving force behind those efforts, what should NIST's role be (especially related to standards), and who should form the backbone of such a collaboration.
Building a successful database has to start with an initial agreement that will provide a guideline for the collaborative effort. 
If provided with clear guidelines, the researchers are willing to share their data, start to collect additional metadata, and standardize their data storing procedures.
With centralized guidelines and a platform where it is easy for everyone to upload the data, establishing a good dataset in a fairly short amount of time -- a big step forward for the quantum dot community -- is within reach.

The Workshop on Advances in Automation of Quantum Dot Devices Control serves as a starting point for discussions about establishing a working group to create a guideline on how to build the centralized database.
The core recommendation here is: 
\begin{enumerate}
    \item The community should establish a working group to oversee the governance and licensing of datasets and to inform database maintainers' choices. 
    NIST could act as a convener of this working group. 
\end{enumerate}

%

\begin{acknowledgments}
This report originated with the \textit{Advances in Automation of Quantum Dot Devices Control} workshop held at NIST on July 19-20, 2023. 
The views and conclusions contained in this paper are those of the authors and should not be interpreted as representing the official policies, either expressed or implied, of the U.S. Government.
The U.S. Government is authorized to reproduce and distribute reprints for Government purposes notwithstanding any copyright noted herein. 
Any mention of equipment, instruments, software, or materials does not imply recommendation or endorsement by the National Institute of Standards and Technology.
\end{acknowledgments}

\section*{Author contributions statement}
JPZ and JMT secured funding and organized the Workshop, coordinated, and contributed to writing the manuscript. 
JPZ, JMT, RA, GB, AC, MAE, EG, MJG, AS, and BW contributed to the scientific content of the meeting and participated in preparing the manuscript.
All other authors contributed to the scientific content of the meeting and provided feedback about the manuscript.

\section*{Competing interests}
The authors declare no competing interests.

\section*{Data availability}
No datasets were analyzed or generated as part of this work. 
All relevant materials can be obtained from the references.

\end{document}